\documentclass[runningheads]{llncs}
\usepackage{amssymb}
\usepackage{multirow}
\usepackage{graphicx}
\usepackage{amsmath}
\usepackage[table]{xcolor}
\usepackage[hidelinks]{hyperref}
\usepackage{fontawesome5}
\usepackage[T1]{fontenc}
\newcommand{\myparagraph}[1]{\smallskip\noindent\textbf{#1}}
\usepackage{enumitem}

\begin{document}
\title{Semi-Supervised Contrastive VAE for Disentanglement of Digital Pathology Images}
%
\titlerunning{SS-cVAE}
%
\author{Mahmudul Hasan\thanks{Email: mahhasan@cs.stonybrook.edu.}\inst{1}
\and
Xiaoling Hu\inst{2} \and 
Shahira Abousamra \inst{1} \and 
Prateek Prasanna \inst{1} \and 
Joel Saltz\inst{1} \and 
Chao Chen\inst{1}} 
\authorrunning{Hasan et al.}
\institute{Stony Brook University, Stony Brook, NY, USA\\
\and
Harvard Medical School, Boston, MA, USA\\
}

\maketitle              
\begin{abstract}
Despite the strong prediction power of deep learning models, their interpretability remains an important concern. Disentanglement models increase interpretability by decomposing the latent space into interpretable subspaces. In this paper, we propose the first disentanglement method for pathology images. We focus on the task of detecting tumor-infiltrating lymphocytes (TIL). We propose different ideas including cascading disentanglement, novel architecture, and reconstruction branches. We achieve superior performance on complex pathology images, thus improving the interpretability and even generalization power of TIL detection deep learning models. Our codes are available at \url{https://github.com/Shauqi/SS-cVAE}.
\end{abstract}

\section{Introduction}
Although deep learning models have strong predictive power, their interpretability remains a significant concern. This is especially important in the medical domain to ensure trust for clinicians and patients. Disentanglement models \cite{higgins2016beta,kim2018disentangling,chen2018isolating} increase interpretability through an encoder-decoder framework. By applying additional regulations such as total correlation to the latent space, one can obtain disentangled latent dimensions corresponding to different interpretable factors, which can be the grounding for training interpretable classifiers \cite{soelistyo2024discovering}.

However, the power of these unsupervised disentanglement methods \cite{higgins2016beta,kim2018disentangling,chen2018isolating} are restricted to relatively simple images, e.g., images with a single object, and will fail to provide satisfying disentanglement for complex images with multiple objects. Recently, a new category of disentanglement methods called contrastive analysis has been proposed \cite{abid2019contrastive,weinberger2022moment,carton2024double}. These methods focus on a discriminative setting. Instead of trying to disentangle the whole latent space, they decompose the latent space into the group of features differentiating different populations, and the group of features that is common across populations. 

In this paper, we propose the first contrastive analysis method for disentangling complex digital pathology images. In particular, we focus on the task of detecting Tumor-Infiltrating Lymphocytes (TILs)~\cite{m_Abousamra_D_F_2022}. TILs are essential in cancer diagnosis and prognosis. Clinical studies have shown that the density of TILs is correlated with overall survival and the length of disease-free survival in multiple cancer types ~\cite{mlecnik2011tumor,angell2013immune,loi:TIL:dfs:2019}. Developing interpretable models will improve transparency and thus trustworthiness of AI methods, facilitating their deployment in clinical scenarios. Furthermore, as we will demonstrate empirically, these interpretable features also have better generalization power. 

Contrastive analysis for TIL prediction is not trivial. Pathology images are highly diverse in tissue types and contain different types of cells, which exhibit a large variation of density and spatial configurations. As demonstrated in Tab.~\ref{Disentanglement Evaluation}, the direct application of the existing contrastive analysis method does not work. To address the challenge, we propose a novel cascaded contrastive analysis method named \emph{Semi-Supervised Contrastive VAE (SS-cVAE)}, consisting of the following technical contributions: 

\begin{enumerate}[topsep=2pt,itemsep=1pt,partopsep=1pt, parsep=1pt]
\item We introduce a cascaded disentangling approach that first differentiates background (tissue) patches and patches with cells and then differentiates patches with low TIL counts and patches with many TILs. This novel cascaded approach proves powerful in disentangling the complex cell spatial configurations in tumor microenvironment.
\item Furthermore, to handle the complex pathology images, we propose substantial architectural improvements from our baseline model MM-cVAE \cite{weinberger2022moment}, including residual blocks in encoders and decoders, as well as three reconstruction branches: the salient reconstruction branch, background reconstruction branch, and classification branch – collectively enhancing the disentanglement performance. 
\item To further improve reconstruction performance (thus the quality of the latent space), we introduce a novel ID-GAN~\cite{lee2020high} reconstruction module. 
\end{enumerate}

\noindent Empirical results demonstrate superior disentangling performance of our method, compared with SOTA approaches. We also show that the disentangled latent features provide strong generalization power across different datasets.

\section{Related Works}
\subsubsection{Unsupervised disentangling methods.}
Existing unsupervised disentanglement approaches like $\beta$-VAE~\cite{higgins2016beta}, Factor VAE~\cite{kim2018disentangling}, $\beta$-TCVAE~\cite{chen2018isolating} are designed to disentangle low-level concepts like position, shape, color, orientation, style, etc. These methods are mostly applied to simple image datasets like dSprites~\cite{dsprites17}, MNIST~\cite{deng2012mnist}, or face image datasets such as CelebA~\cite{liu2015faceattributes}. They work when the image only contains a single object/face. When applied to digital pathology images, these unsupervised disentangling VAEs are limited to disentangling factors like hue and saturation and fail to capture the complex spatial relationship between cells. Additionally, their reconstruction performance often results in significant blurriness, making it challenging to interpret the disentangled factors from these models. The unique challenge presented by digital pathology images in disentanglement tasks involves multiple concepts, including spatial and morphological variations in diverse objects, as well as variations in tissue types, posing a much more complex scenario than what these methods can handle.

\subsubsection{Contrastive analysis based disentangling methods.}
In contrast to unsupervised disentanglement methods, contrastive analysis approaches incorporate supervision using classifiers to decompose the latent space dimensions into two groups. One group contains latent features shared by different classes. Whereas the other group contains latent features differentiating different classes. Existing methods include Contrastive Analysis VAE (CA-VAE) methods (including cVAE~\cite{abid2019contrastive}, MM-cVAE~\cite{weinberger2022moment}) and CA-GAN based methods (such as Double InfoGan~\cite{carton2024double}). These methods have demonstrated their ability to disentangle on datasets such as Gene Expression~\cite{haber2017single,zheng2017massively}, CIFAR-10~\cite{krizhevsky2009learning}, MNIST~\cite{deng2012mnist}, and medical image datasets like BRATS~\cite{menze2014multimodal}. However, they are primarily focusing on single-object images. When applied to complex pathology images containing multiple objects having complex spatial or topological relationships, the disentanglement and reconstruction performance is unsatisfactory. 

\begin{figure}[t]
    \centering
    \includegraphics[width=0.85\textwidth]{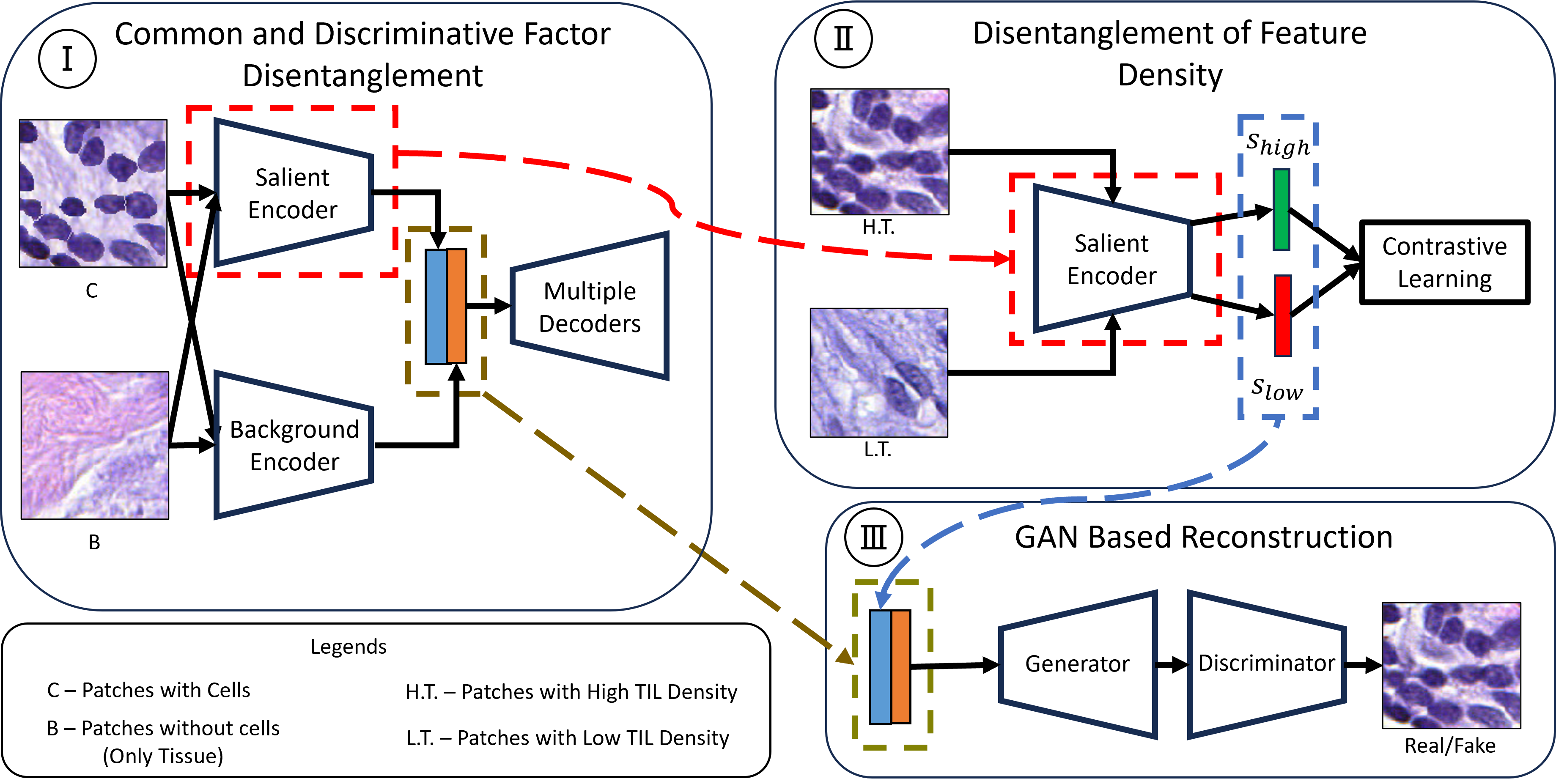}
    \caption{Whole SS-cVAE Framework. I) Module to disentangle discriminative factors from common factors. II) Module for disentangling factors related to density. Note that, the same salient encoder from (I) is fine-tuned here. III) Module for image reconstruction from the disentangled latent generated from (I) and (II).}
    \label{Semi-Supervised Contrastive VAE Framework}
\end{figure}

\section{Method}
Our disentanglement framework, illustrated in Fig.~\ref{Semi-Supervised Contrastive VAE Framework}, is comprised of three modules. 
In the first module (Fig.~\ref{Semi-Supervised Contrastive VAE Framework}(I)), disentanglement occurs on synthetic patches containing cells, $C$, and real tissue-only patches, $B$. Utilizing a copy-paste mechanism in synthetic data creation, pairwise patches from both $C$ and $B$ are sent to two encoders: a salient encoder and a background encoder. These encoders separate patch embeddings into salient latent space (the discriminative factor, i.e. the cells) and background latent space (the common factor, i.e. the tissue). The use of pairwise synthetic images enhances disentanglement compared to unpaired real images, aiding the model in understanding the appearance with and without the discriminating factor. 

In the second module (Fig.~\ref{Semi-Supervised Contrastive VAE Framework}(II)), the salient encoder is fine-tuned to disentangle the density of a specific type of object - lymphocytes in our case. The rationale behind the two-stage disentanglement is to not only capture the existence of lymphocytes but also the variation in lymphocyte density. A two-stage disentanglement makes disentangling complex factors easier, so each module focuses on one pair of populations (with vs.~without cells, low vs.~high lymphocytes).
Compressing the disentanglement into a single step could result in noisy latent factorization involving variations in other cell types and tissue characteristics. The two-stage disentanglement addresses this issue by eliminating tissue variation in the initial stage, leaving only other cell variations to be considered. 

The third module is an image reconstruction model. Given the observed poor reconstruction performance of the VAE decoder, a GAN-based module is introduced (Fig.~\ref{Semi-Supervised Contrastive VAE Framework}(III)). This module takes as input the latent embeddings generated by the salient and background encoders and attempts to reconstruct the original image. Further details about each module are provided in the subsequent sections. 

\myparagraph{Module 1: common and discriminative factor disentanglement.}
\begin{figure}[t]
    \centering
    \includegraphics[width=0.70\textwidth]{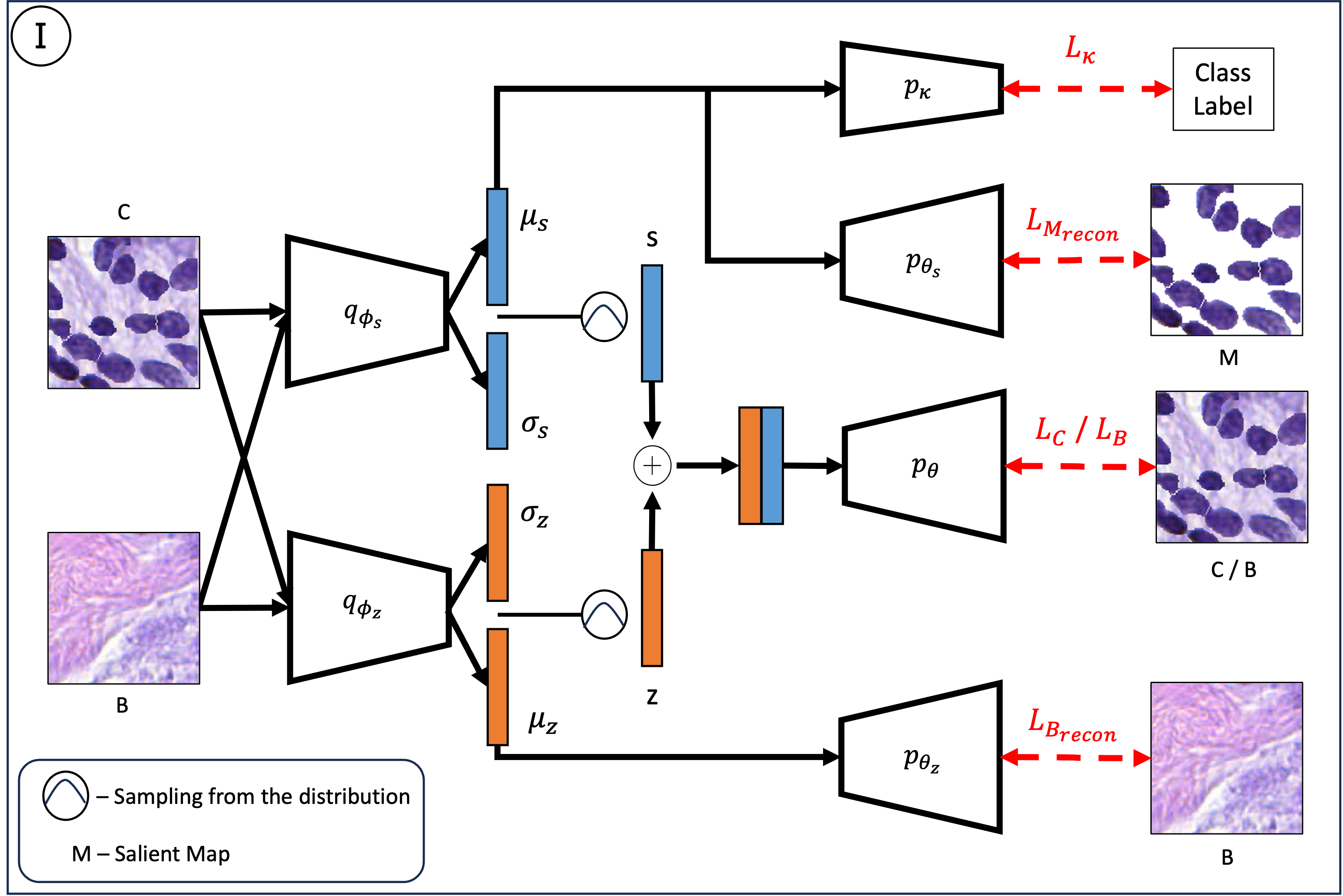}
    \caption{The first module of SS-cVAE. It performs disentanglement of discriminative factor from common factor given any patch of $C$ or $B$.}
    \label{Semi-Supervised Contrastive VAE Architecture for Common and Discriminative Factor Disentanglement}
\end{figure}
Fig.~\ref{Semi-Supervised Contrastive VAE Architecture for Common and Discriminative Factor Disentanglement} illustrates the architecture of this disentanglement module for $C$ and $B$. It comprises two probabilistic encoders, $q_{\phi_s}$ and $q_{\phi_z}$ approximating the salient likelihood $\mathcal{N}(\mu_s, \sigma_s)$ and background likelihood $\mathcal{N}(\mu_z,\sigma_z)$, respectively. Salient latent code, $s$, and background latent code, $z$, are sampled from these distributions. The network is trained with three reconstruction branches: image reconstructor, $P_\theta$, salient reconstructor, $P_{\theta_s}$, and background reconstructor, $P_{\theta_z}$. We also have a classification branch, $P_\kappa$. $P_\theta$ reconstructs the original image from the concatenation of sampled latent encodings $s$ and $z$. $P_{\theta_s}$ reconstructs the salient map, $M$ from $\mu_s$. $P_{\theta_z}$ reconstructs the background image $B$ from $\mu_z$.  The module is trained with the following objective function:
\begin{equation}
\label{Eq6}
\begin{aligned}
     \mathcal{L}_c + \mathcal{L}_b + \mathcal{L}_{M_{recon}} + \mathcal{L}_{B_{recon}} + \mathcal{L}_\kappa
\end{aligned}
\end{equation}
where $\mathcal{L}_c$ is the Evidence Lower Bound (ELBO) for $C$,  $\mathcal{L}_b$ is the ELBO for $B$,
$\mathcal{L}_{M_{recon}}$ is the loss for salient reconstruction, $\mathcal{L}_{B_{recon}}$ is the loss for background reconstruction, and $\mathcal{L}_\kappa$ is the binary cross entropy loss for classifying whether a patch contains cells or not. For the reconstructors, we use mean squared error (MSE) to calculate the losses. The equations for the loss functions are in Eq.~3-6 in the supplementary materials.

\myparagraph{Module 2: disentanglement of feature density.}
\begin{figure}[t]
    \centering
    \includegraphics[width=0.85\textwidth]{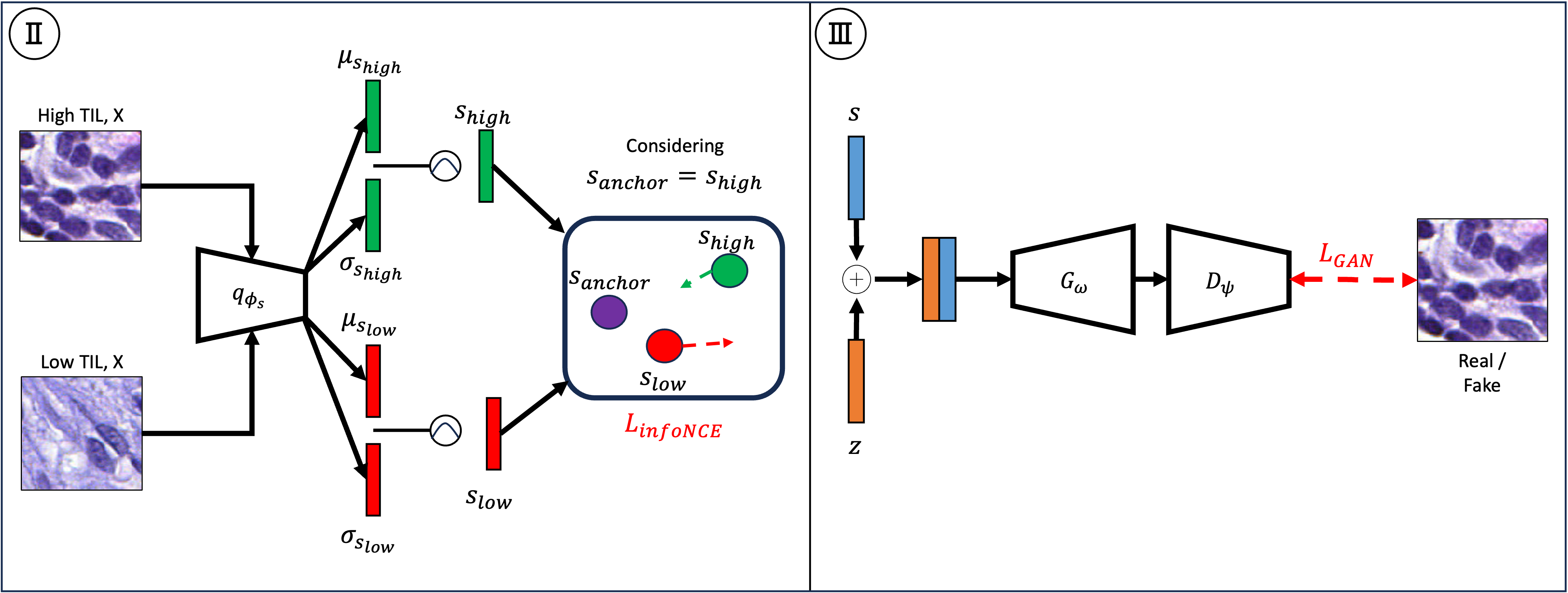}
    \caption{2nd and 3rd modules of SS-cVAE. II) Module to disentangle the factor corresponding to density. During InfoNCE loss calculation in (II) we consider one of the $s_{high}$ samples as $s_{anchor}$  III) GAN-based reconstruction using disentangled latent.}
    \label{Semi-Supervised Contrastive VAE High TIL vs Low TIL Disentanglement and Reconstruction}
\end{figure}
This module focuses on creating an embedding that discerns varying densities of a cell-containing image, such as the presence level of a specific cell type in a patch. To achieve this, we fine-tune the previously trained salient encoder, resulting in a salient latent distribution $\mathcal{N}(\mu_s,\sigma_s)$ that captures feature density. During training, latent encodings $s_{high}$ and $s_{low}$ are sampled from patches with high and low feature densities, respectively. InfoNCE loss is applied to $s_{high}$ and $s_{low}$ to push the two distributions apart. Additionally, KL-Divergence loss is incorporated to fine-tune the distribution itself. Fig.~\ref{Semi-Supervised Contrastive VAE High TIL vs Low TIL Disentanglement and Reconstruction}(II) illustrates this module. The training objective function is as follows:
\begin{equation}
\label{Objective for Density}
    \mathcal{L}_{infoNCE} - \mathcal{L}_{KL_s}
\end{equation}
The equations for $\mathcal{L}_{infoNCE}$ and $\mathcal{L}_{KL_s}$ are in Eq.~7-8 in supplementary material.

\myparagraph{Module 3: GAN-based reconstruction.}
While VAEs serve as effective latent distribution encoders, their image generation quality can be suboptimal. To enhance image reconstruction, we employ a GAN-based model, specifically IDGAN~\cite{lee2020high}. Training IDGAN involves reconstructing images from the concatenation of disentangled salient and background embeddings, yielding superior reconstruction compared to the VAE decoder. Fig.~\ref{Semi-Supervised Contrastive VAE High TIL vs Low TIL Disentanglement and Reconstruction}(III) presents the conceptual diagram of IDGAN. The loss function for IDGAN, $\mathcal{L}_{GAN}(D, G)$, is detailed in Eq.~9 in the supplementary materials.

\section{Experiments and Results}
\myparagraph{Datasets and preprocessing.}
\begin{figure}[t]
    \centering
    \includegraphics[width=0.8\textwidth]{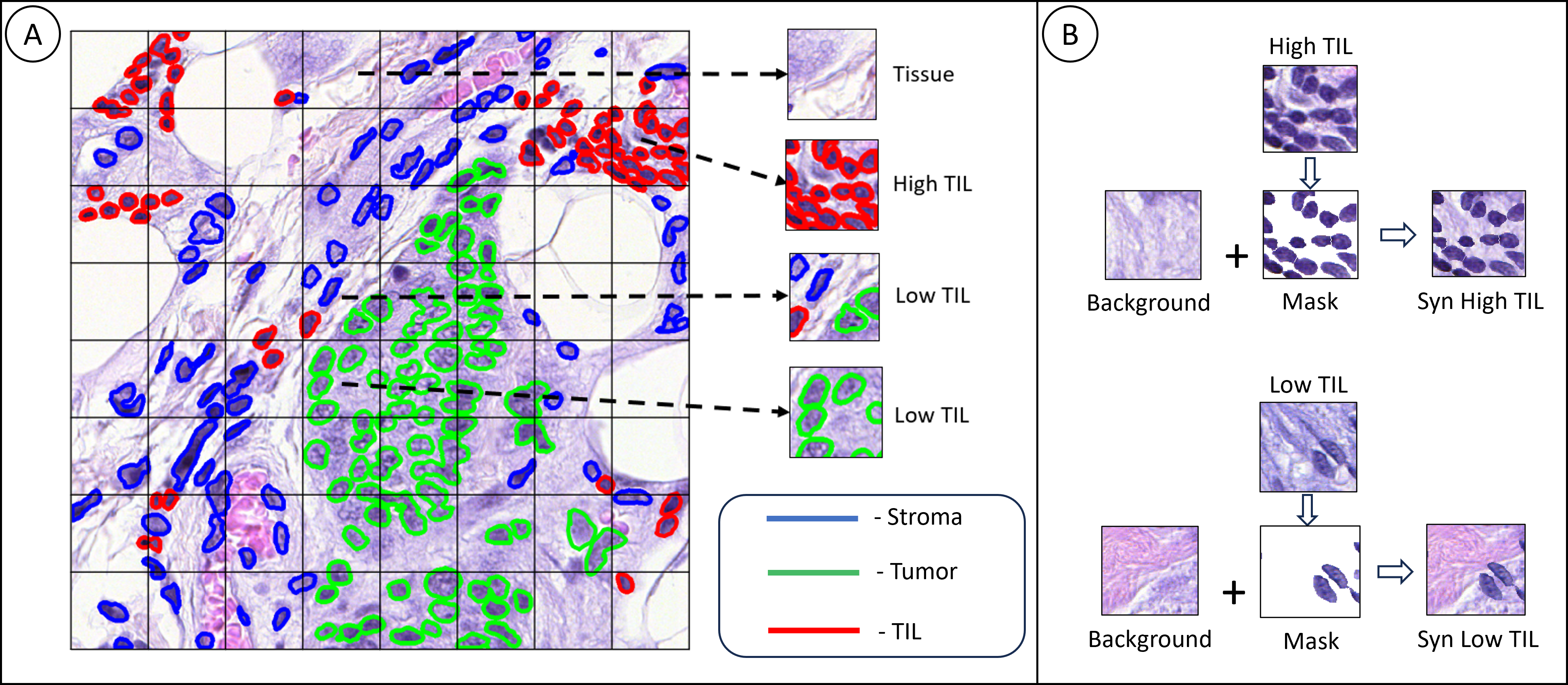}
    \caption{Data Preprocessing. A) Patch extraction and patch labeling using ground truth cell annotation. B) Synthetic Data Creation using a copy-paste mechanism.}
    \label{Patch Extraction and Labeling}
\end{figure}
For the disentanglement task, synthetic data is generated using the TCGA BRCA dataset~\cite{abousamra2021multi} and the CoNSeP dataset~\cite{graham2019hover}. Both datasets consist of $1000 \times 1000$ pixel tiles in native magnification close to 40$\times$. The data statistics are in Tab.~4 of the supplementary materials. Fig.~\ref{Patch Extraction and Labeling}(A) illustrates the data preprocessing, including the extraction and labeling of training patches. Tiles are resized to $1024\times1024$ images, and non-overlapping $128\times128$ patches are extracted. Both datasets contain ground truth cell annotations with their corresponding class. Using these annotations we assign patch-level labels based on the frequency of cells. For the first-stage disentanglement, patches are divided into two groups: \textit{patches with cells} ($C$) and \textit{patches without cells} ($B$). The cell-to-patch area ratio determines the grouping, with patches having a ratio greater than 10\% assigned to $C$; the remaining patches are grouped into $B$. This partition is balanced (1:1) across training, validation, and test sets (Tab.~4, Supp.). For the second-stage disentanglement, patches are labeled as \textit{low TIL density} ($L.T.$) or \textit{high TIL density} ($H.T.$) based on TIL instances. Patches with at least 5 TIL instances are categorized as $H.T.$, while others are labeled as $L.T.$ (Tab.~4, Supp.). The $C$ and $B$ disentanglement model is trained on a synthetic dataset created using a copy-paste mechanism. A random cell mask from $C$ patches is copied and pasted onto random $B$ patches (Fig.~\ref{Patch Extraction and Labeling}(B)). For downstream tasks, three classes are considered: patches with only tissue, patches with only TILs, and patches with other cells (i.e., stroma, tumor). Statistics for the downstream task data are detailed in Tab.~4 (Supp.).

\myparagraph{Implementation details.}
Our model employs residual blocks in encoders and transposed residual blocks in decoders. A batch size of 64 is consistently used for training, validation, and test set loading across all models, including baselines. The Adam optimizer is employed with a learning rate of 0.001. Hyperparameters in Eq.~3 and Eq.~4 (refer to Supplementary Materials) are configured as $\lambda_1 = 10$ and $\lambda_2 = 10$. For both $\lambda_1 = 10$ and $\lambda_2 = 10$, we tested different values ranging from $10^{-3}$ to $10^3$ with steps of multiples of 10 on the BRCA dataset and selected the one with the best validation performance. The sizes of the salient and background latent vectors are set to $s = 128$ and $z = 128$.

\myparagraph{Evaluation metrics.}
For evaluation, the silhouette score (SS) is employed, as described in~\cite{carton2024double} and~\cite{weinberger2022moment}, on a real test set. Additionally, the mean of the latent is computed, augmented with a simple linear classifier, and subjected to a 5-fold cross-validation binary classification on the test set. Reconstruction performance is assessed using Frechet Inception Distance (FID). Lastly, for downstream task evaluation, multiclass classification scores are utilized.

\myparagraph{Quantitative evaluations.}
Tab.~\ref{Disentanglement Evaluation} presents the disentanglement evaluation of our model against SOTA baselines. A higher value for the salient latent indicates superior encoding of discriminative factors, while a lower value for the background latent suggests an effective representation of a common factor without discriminative power. Our model excels in salient disentanglement, outperforming baselines, and demonstrates comparable performance in background latent disentanglement. For High TIL and Low TIL density, our model surpasses the baselines, too.
\begin{table}[t]
\centering
\caption{Disentanglement Evaluation. The models are trained on synthetic datasets and evaluated against real datasets. B = Patches without cells, C = Patches with cells, H.T. = High TIL, L.T. = Low TIL, SS = Silhouette Score, S = Salient Latent Space, Z = Background Latent Space, CLF = Classification Score}
\label{Disentanglement Evaluation}
\begin{tabular}{ccccccc}
\hline
Model & \multicolumn{2}{c}{B vs C} & H.T. vs L.T. & \multicolumn{2}{c}{B vs C} & H.T. vs L.T.\\
& $SS_S \uparrow$  & $SS_Z \downarrow$ & $SS_S \uparrow$ & $CLF_S \uparrow$  & $CLF_Z \downarrow$ & $CLF_S \uparrow$ \\
\hline
\multicolumn{7}{c}{Dataset: BRCA}\\
\hline
$\beta$-VAE & 0.05 & 0.01 & - & 0.64$\pm$0.07 & 0.49$\pm$0.08 & - \\
Factor-VAE & 0.06 & 0.07 & - & 0.61$\pm$0.06 & 0.60$\pm$0.08 & -\\
$\beta$-TCVAE & 0.13 & 0.08 & - & 0.70$\pm$0.06 & 0.62$\pm$0.10 & -\\
Double InfoGAN & 0.04 & 0.01 & 0.005 & 0.83$\pm$0.07 & 0.70$\pm$0.06& 0.65$\pm$0.05\\
cVAE & 0.10 & 0.20 & 0.10 & 0.89$\pm$0.03 & 0.92$\pm$0.06 & 0.76$\pm$0.02\\
MM-cVAE & 0.14 & 0.13 & 0.06 & 0.86$\pm$0.04 & 0.92$\pm$0.03 & 0.75$\pm$0.04\\
\hline
SS-cVAE & 0.32 & 0.10 & 0.27 & 0.92$\pm$0.04 & 0.92$\pm$0.02 & 0.87$\pm$0.05\\
\hline
\multicolumn{7}{c}{Dataset: CoNSeP}\\
\hline
$\beta$-VAE & 0.08 & 0.007 & - & 0.65$\pm$0.09 & 0.53$\pm$0.06 & -\\
Factor-VAE & 0.009 & 0.007 & - & 0.57$\pm$0.08 & 0.53$\pm$0.02 & -\\
$\beta$-TCVAE & 0.10 & 0.006 & - & 0.66$\pm$0.04 & 0.57$\pm$0.09 & -\\
Double InfoGAN & 0.006 & 0.009 & 0.005 & 0.64$\pm$0.07 & 0.55$\pm$0.06&0.57$\pm$0.09\\
cVAE & 0.16 & 0.12 & 0.04 & 0.66$\pm$0.10 & 0.62$\pm$0.08 & 0.67$\pm$0.13 \\
MM-cVAE & 0.06 & 0.03 & 0.11 & 0.64$\pm$0.12& 0.64$\pm$0.06 & 0.69$\pm$0.06\\
\hline
SS-cVAE & 0.18 & 0.04 & 0.13 & 0.85$\pm$0.02 & 0.75$\pm$0.07 & 0.78$\pm$0.09\\
\hline
\end{tabular}
\end{table}
Reconstruction evaluation against contrastive analysis (CA) baselines is detailed in Tab.~\ref{Reconstruction and Downstream}. The results indicate that our model's reconstruction performance is superior to baselines, except for Double InfoGan. Importantly, our model achieves better disentanglement than Double InfoGan. 
\begin{table}[t]
    \centering
    \caption{Reconstruction Evaluation and Downstream Task Evaluation}
    \label{Reconstruction and Downstream}
    \begin{tabular}{cc|cc}
    \hline
    \multicolumn{2}{c|}{Reconstruction} & \multicolumn{2}{c}{Downstream 3 Class Classification on CoNSeP dataset}\\
    \hline
    Model & FID & Model & Accuracy\\
    \hline
    MM-cVAE & 293.71 & ResNet-18 (without pre-train) & 0.71\\
    Double InfoGAN & 214.59 & ResNet-18 (BRCA pretrain) & 0.76\\
    SS-cVAE & 221.15 & SS-cVAE (BRCA pretrain) & 0.79\\
    \hline
    \end{tabular}
\end{table}
Finally, utilizing our model as a pre-trained feature extractor on the BRCA dataset and finetuning on the CoNSeP dataset for multiclass classification as a downstream task, we compare our model with ResNet18 in Tab.~\ref{Reconstruction and Downstream}. For the downstream task, we have used 3 classes: patches with only tissue, patches with only TILs, and patches with other cells (i.e., stroma, tumor).  The findings suggest that our model can outperform the ResNet18 model in multiclass classification within the transfer learning framework. Apart from the abovementioned experiments we also present ablation experiments with different architectural changes and reconstructions in Tab.~3 (Supp.).

\begin{figure}[t]
    \centering
    \includegraphics[width=0.85\textwidth]{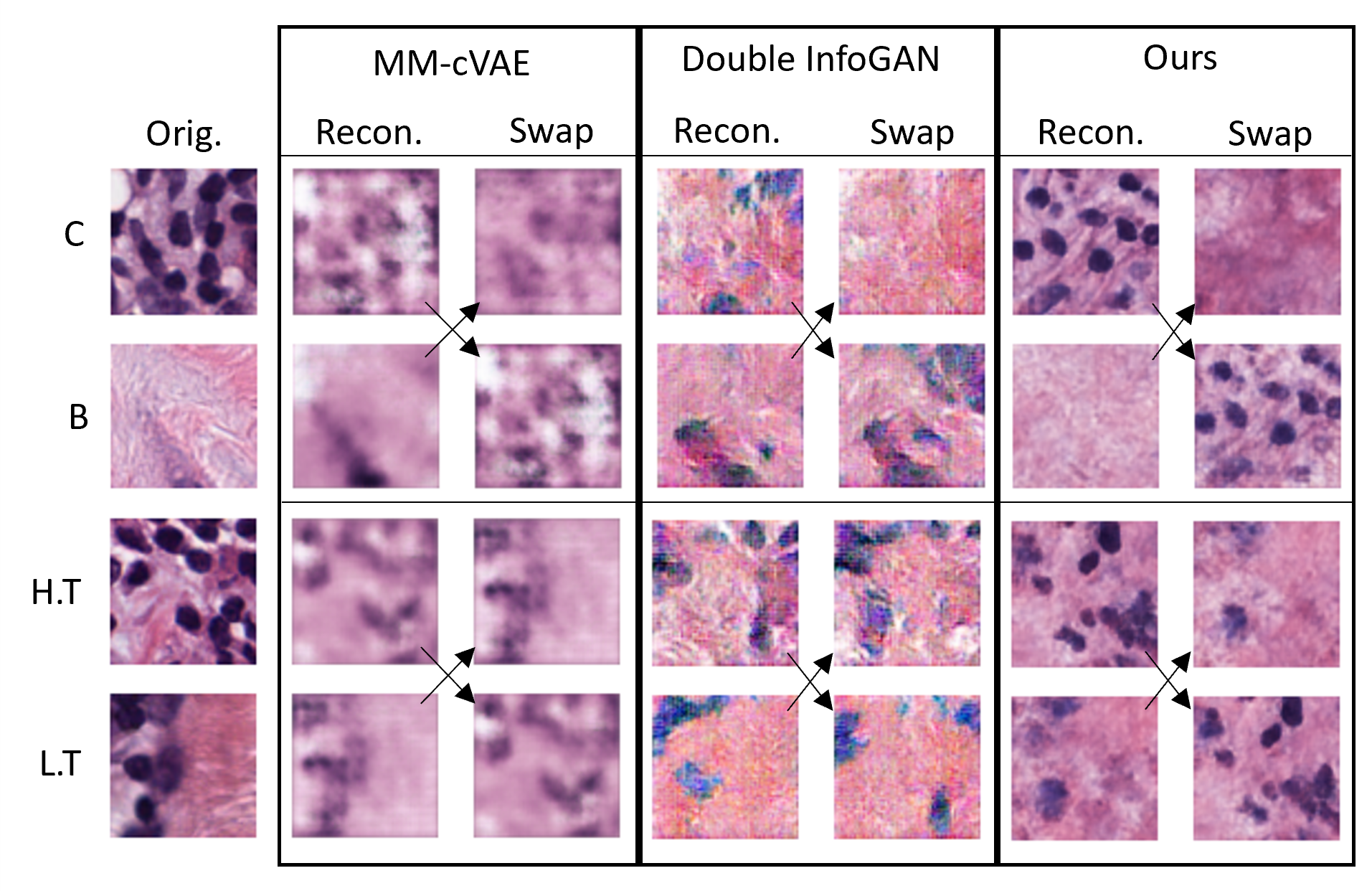}
    \caption{Example of Latent Swap. Row 1 and 2 represent salient latent swapping between the samples of $C$ and $B$. Row 3 and 4 represent the salient latent swapping between the samples of $H.T$ and $L.T$.}
    \label{Latent Swap}
\end{figure}

\myparagraph{Qualitative evaluations.}
To visually depict what is encapsulated in the salient and background latent spaces, we conduct latent swapping between two samples. Fig.~\ref{Latent Swap} exhibits salient latent swaps between patches from $C$ and $B$ (rows 1 and 2) and between $H.T.$ and $L.T$ patches (rows 3 and 4). The results demonstrate that our model's generations accurately capture the content of the original image, encompassing both content and cell/TIL density. In contrast, other baselines exhibit complete failure in this regard. This emphasizes the superior quality of the embeddings generated by our proposed SS-cVAE. Note that we omit results for $\beta$-VAE, Factor VAE, and $\beta$-TCVAE as they produce highly blurry and non-interpretable reconstructions.
Additionally, we explore latent space interpolation through Fig.~6 in the supplementary materials. This demonstrates synthetic image generation via latent space traversal/interpolation between tissue/background images and cell images, as well as between low and high TIL images. This experimentation showcases our ability to synthesize new images while controlling discriminating factors, common factors, and density factors.
\section{Conclusions}
In this paper, we introduced a novel framework for disentangling concepts, including discriminative, common, and density factors, within distinct populations of digital pathology samples. Our proposed framework demonstrated notable enhancements, showcasing substantial qualitative and quantitative improvements over existing approaches in disentanglement tasks.

\subsubsection{\ackname}
This research was partially supported by the National Science Foundation (NSF) grant CCF-2144901, the National Institute of General Medical Sciences (NIGMS) grant R01GM148970, the National Institute of Health (NIH) grant 5R21CA258493-02, the National Cancer Institute (NCI) grant UH3-CA225021 and the Stony Brook Trustees Faculty Award. This work used Bridges-2 \cite{brown2021bridges} at Pittsburgh Supercomputing Center through allocation [asc13007p] from the Advanced Cyberinfrastructure Coordination Ecosystem: Services \& Support (ACCESS) program, which is supported by NSF grants.

\subsubsection{Disclosure of Interests.}
The authors have no competing interests to declare that are relevant to the content of this article. Co-Author Joel Saltz is a co-founder of Chilean Wool LLC.

\bibliographystyle{splncs04}
\bibliography{ref,frontiers_ref}

\begin{thebibliography}{10}
\providecommand{\url}[1]{\texttt{#1}}
\providecommand{\urlprefix}{URL }
\providecommand{\doi}[1]{https://doi.org/#1}

\bibitem{abid2019contrastive}
Abid, A., Zou, J.: Contrastive variational autoencoder enhances salient features. arXiv preprint arXiv:1902.04601  (2019)

\bibitem{abousamra2021multi}
Abousamra, S., Belinsky, D., Van~Arnam, J., Allard, F., Yee, E., Gupta, R., Kurc, T., Samaras, D., Saltz, J., Chen, C.: Multi-class cell detection using spatial context representation. In: ICCV (2021)

\bibitem{m_Abousamra_D_F_2022}
Abousamra, S., Gupta, R., Hou, L., Batiste, R., Zhao, T., Shankar, A., Rao, A., Chen, C., Samaras, D., Kurc, T., Saltz, J.: Deep learning-based mapping of tumor infiltrating lymphocytes in whole slide images of 23 types of cancer. In: Frontiers in oncology. p.~5971 (2022), \url{https://www.frontiersin.org/articles/10.3389/fonc.2021.806603/full}

\bibitem{angell2013immune}
Angell, H., Galon, J.: From the immune contexture to the immunoscore: the role of prognostic and predictive immune markers in cancer. Current opinion in immunology  \textbf{25}(2),  261--267 (2013)

\bibitem{brown2021bridges}
Brown, S.T., Buitrago, P., Hanna, E., Sanielevici, S., Scibek, R., Nystrom, N.A.: Bridges-2: A platform for rapidly-evolving and data intensive research. In: Practice and Experience in Advanced Research Computing (2021)

\bibitem{carton2024double}
Carton, F., Louiset, R., Gori, P.: Double infogan for contrastive analysis. arXiv preprint arXiv:2401.17776  (2024)

\bibitem{chen2018isolating}
Chen, R.T., Li, X., Grosse, R.B., Duvenaud, D.K.: Isolating sources of disentanglement in variational autoencoders. In: NeurIPS (2018)

\bibitem{deng2012mnist}
Deng, L.: The mnist database of handwritten digit images for machine learning research. IEEE Signal Processing Magazine  (2012)

\bibitem{graham2019hover}
Graham, S., Vu, Q.D., Raza, S.E.A., Azam, A., Tsang, Y.W., Kwak, J.T., Rajpoot, N.: Hover-net: Simultaneous segmentation and classification of nuclei in multi-tissue histology images. Medical image analysis  (2019)

\bibitem{haber2017single}
Haber, A.L., Biton, M., Rogel, N., Herbst, R.H., Shekhar, K., Smillie, C., Burgin, G., Delorey, T.M., Howitt, M.R., Katz, Y., et~al.: A single-cell survey of the small intestinal epithelium. Nature  (2017)

\bibitem{higgins2016beta}
Higgins, I., Matthey, L., Pal, A., Burgess, C., Glorot, X., Botvinick, M., Mohamed, S., Lerchner, A.: beta-vae: Learning basic visual concepts with a constrained variational framework. In: ICLR (2016)

\bibitem{kim2018disentangling}
Kim, H., Mnih, A.: Disentangling by factorising. In: ICML (2018)

\bibitem{krizhevsky2009learning}
Krizhevsky, A., Hinton, G., et~al.: Learning multiple layers of features from tiny images  (2009)

\bibitem{lee2020high}
Lee, W., Kim, D., Hong, S., Lee, H.: High-fidelity synthesis with disentangled representation. In: ECCV (2020)

\bibitem{liu2015faceattributes}
Liu, Z., Luo, P., Wang, X., Tang, X.: Deep learning face attributes in the wild. In: ICCV (2015)

\bibitem{loi:TIL:dfs:2019}
Loi, S., Drubay, D., Adams, S., Pruneri, G., Francis, P., Lacroix-Triki, M., Joensuu, H., Dieci, M., Badve, S., Demaria, S., Gray, R., Munzone, E., Lemonnier, J., Sotiriou, C., Piccart, M., Kellokumpu-Lehtinen, P.L., Vingiani, A., Gray, K., Andre, F., Michiels, S.: Tumor-infiltrating lymphocytes and prognosis: A pooled individual patient analysis of early-stage triple-negative breast cancers. Journal of Clinical Oncology  \textbf{37},  JCO.18.01010 (01 2019). \doi{10.1200/JCO.18.01010}

\bibitem{dsprites17}
Matthey, L., Higgins, I., Hassabis, D., Lerchner, A.: dsprites: Disentanglement testing sprites dataset. https://github.com/deepmind/dsprites-dataset/ (2017)

\bibitem{menze2014multimodal}
Menze, B.H., Jakab, A., Bauer, S., Kalpathy-Cramer, J., Farahani, K., Kirby, J., Burren, Y., Porz, N., Slotboom, J., Wiest, R., et~al.: The multimodal brain tumor image segmentation benchmark (brats). IEEE transactions on medical imaging  (2014)

\bibitem{mlecnik2011tumor}
Mlecnik, B., Bindea, G., Pag{\`e}s, F., Galon, J.: Tumor immunosurveillance in human cancers. Cancer and Metastasis Reviews  \textbf{30}(1),  5--12 (2011)

\bibitem{soelistyo2024discovering}
Soelistyo, C.J., Lowe, A.R.: Discovering interpretable models of scientific image data with deep learning. arXiv preprint arXiv:2402.03115  (2024)

\bibitem{weinberger2022moment}
Weinberger, E., Beebe-Wang, N., Lee, S.I.: Moment matching deep contrastive latent variable models. arXiv preprint arXiv:2202.10560  (2022)

\bibitem{zheng2017massively}
Zheng, G.X., Terry, J.M., Belgrader, P., Ryvkin, P., Bent, Z.W., Wilson, R., Ziraldo, S.B., Wheeler, T.D., McDermott, G.P., Zhu, J., et~al.: Massively parallel digital transcriptional profiling of single cells. Nature communications  (2017)

\end{thebibliography}
\end{document}


\title{Semi-Supervised Contrastive VAE for Disentanglement of Digital Pathology Images \\ --- Supplementary Material ---}
\titlerunning{SS-cVAE}
\author{}
\institute{}

\maketitle              

\setcounter{figure}{5}
\setcounter{table}{2}
\setcounter{equation}{2}

\begin{equation}
\label{L_b Loss}
\begin{aligned}
\mathcal{L}_b(B) &= \sum_{j=1}^m(\mathop{{}\mathbb{E}}_{q_{\phi_z}(z)}[\log{p_\theta}(B_j|z,s')] - D_{KL}(q_{\phi_z}(z|B_j)||p_b(z))) \\
&- \lambda_1 \cdot L_{MMD}(\hat{q}_{\phi_{s,b}}(s),\delta\{s=s'\})
\end{aligned}
\end{equation}

\begin{equation}
\label{L_c Loss}
\begin{aligned}
    \mathcal{L}_c(C_i) &= \mathop{{}\mathbb{E}}_{q_{\phi_z}(z)q_{\phi_s}(s)}[\log{p_\theta}(C_i|z,s)] - D_{KL}(q_{\phi_z}(z|C_i)||p_c(z)) \\
    & - D_{KL}(q_{\phi_s}(s|C_i)||p_c(s)) - \lambda_2 \mathcal{L}_{MMD}(\hat{q}_{\phi_{z,x}}(z), \hat{q}_{\phi_{z,b}}(z))
\end{aligned}
\end{equation}

\begin{equation}
\label{L_M_recon}
    \mathcal{L}_{M_{recon}} = \mathop{{}\mathbb{E}}_{q_{\phi_s}(s)}[\log{p_{\theta_s}}(M|\mu_s)]
\end{equation}

\begin{equation}
\label{L_B_recon}
    \mathcal{L}_{B_{recon}} = \mathop{{}\mathbb{E}}_{q_{\phi_z}(z)}[\log{p_{\theta_z}}(B|\mu_z)]
\end{equation}

\begin{equation}
\label{L_InfoNCE}
\begin{aligned}
    \mathcal{L}_{InfoNCE} = -log[\frac{exp(v \cdot v^+ / \tau)}{exp(v \cdot v^+ / \tau) + \sum_{n=1}^N exp(v \cdot v_n^- / \tau)}]
\end{aligned}
\end{equation}

\begin{equation}
\label{L_KL_s}
\begin{aligned}
    \mathcal{L}_{KL_s} 
    = D_{KL}(q_{\phi_s}(s|x_i)||p_x(z))
\end{aligned}
\end{equation}

\begin{equation}
\label{IDGAN_EQ}
\begin{aligned}
    \mathcal{L}_{GAN}(D,G) = \mathop{{}\mathbb{E}}_{x \sim p(x)}[log{D(x)}] + \mathop{{}\mathbb{E}}_{s \sim q_{\phi_s}(s), z \sim q_{\phi_z}(z)}[log(1 - D(G(s,z)))]
\end{aligned}
\end{equation}

\begin{table}[h]
\centering
\caption{Ablation Study of disentanglement models and reconstructors on BRCA dataset. B = Patches without cells, C = Patches with cells, H.T. = High TIL, L.T. = Low TIL, SS = Silhouette Score, S = Salient Latent Space, Z = Background Latent Space}
\label{Ablation Study}
\begin{tabular}{ccccc}
\hline
Model & \multicolumn{2}{c}{B vs C} & H.T. vs L.T. & FID\\
& $SS_S \uparrow$  & $SS_Z \downarrow$ & $SS_S \uparrow$ & \\
\hline
MM-cVAE (ResNet Architecture) & 0.28 & 0.13 & 0.11 & -\\
MM-cVAE (ResNet Architecture) + $L_\kappa$ & 0.28 & 0.13 & 0.11 & -\\
MM-cVAE (ResNet Architecture) + $L_{M_{recon}}$ & 0.04 & 0.12 & 0.04 & -\\
MM-cVAE (ResNet Architecture) + $L_{B_{recon}}$ & 0.29 & 0.10 & 0.13 & -\\
SS-cVAE (One Step Disentanglement) & 0.28 & 0.08 & 0.05 & -\\
SS-cVAE with VAE as Reconstructor & - & - & - & 251.32\\
SS-cVAE with IDGAN as Reconstructor & 0.32 & 0.10 & 0.27 & 214.59\\
\hline
\end{tabular}
\end{table}

\begin{table}
\centering
\caption{Dataset Statistics. C = patches with cells, B = patches without cells, H.T. = patches with High TIL density, L.T. = patches with low TIL density}
\label{Dataset Statistics}
\begin{tabular}{c|ccc|cccccc|cccccc}
\hline
\multirow{3}{*}{Dataset} & \multicolumn{3}{c|}{Tiles ($1000\times1000$)} &\multicolumn{6}{c|}{Patches ($128\times128$)} & \multicolumn{6}{c}{Patches ($128\times128$)} \\
&Train & Valid & Test &\multicolumn{2}{c}{Train} & \multicolumn{2}{c}{Valid} & \multicolumn{2}{c|}{Test}&\multicolumn{2}{c}{Train} & \multicolumn{2}{c}{Valid} & \multicolumn{2}{c}{Test}\\
&&&&B&C&B&C&B&C&L.T&H.T&L.T&H.T&L.T&H.T\\
\hline
BRCA & 94 & 9 & 10 & 508 & 508 & 14 & 14 & 78 & 78 & 354 & 354 & 30 & 30 & 115 & 115\\
CoNSeP & 27 & - & 14 & 479 & 479 & 70 & 70 & 70 & 70 & 398 & 398 & 50 & 50 & 55 & 55\\
\hline
\multicolumn{16}{c}{Mean and Std of Cell Count Per Tiles and Per Patches for BRCA dataset}\\
\hline
 & \multicolumn{3}{c|}{Cell count in Tiles} &\multicolumn{6}{c|}{Cell Count in C}&\multicolumn{6}{c}{TIL count in H.T.}\\
&Train & Valid & Test &\multicolumn{2}{c}{Train} & \multicolumn{2}{c}{Valid} & \multicolumn{2}{c|}{Test}&\multicolumn{2}{c}{Train} & \multicolumn{2}{c}{Valid} & \multicolumn{2}{c}{Test}\\
\hline
mean&281&271&355& \multicolumn{2}{c}{6} & \multicolumn{2}{c}{7} & \multicolumn{2}{c|}{7} &\multicolumn{2}{c}{9} & \multicolumn{2}{c}{8} &\multicolumn{2}{c}{11} \\
std&134&54&173&\multicolumn{2}{c}{3}&\multicolumn{2}{c}{3}&\multicolumn{2}{c|}{4} & \multicolumn{2}{c}{3} & \multicolumn{2}{c}{2} &\multicolumn{2}{c}{3}\\
\hline
\multicolumn{16}{c}{Mean and Std of Cell Count Per Tiles and Per Patches for CoNSeP dataset}\\
\hline
 & \multicolumn{3}{c|}{Cell count in Tiles} &\multicolumn{6}{c|}{Cell Count in C}&\multicolumn{6}{c}{TIL count in H.T.}\\
&Train & Valid & Test &\multicolumn{2}{c}{Train} & \multicolumn{2}{c}{Valid} & \multicolumn{2}{c|}{Test}&\multicolumn{2}{c}{Train} & \multicolumn{2}{c}{Valid} & \multicolumn{2}{c}{Test}\\
\hline
mean&576&-&627& \multicolumn{2}{c}{10} & \multicolumn{2}{c}{9} & \multicolumn{2}{c|}{9} &\multicolumn{2}{c}{9} & \multicolumn{2}{c}{10} &\multicolumn{2}{c}{9} \\
std&477&-&325&\multicolumn{2}{c}{6}&\multicolumn{2}{c}{4}&\multicolumn{2}{c|}{6} & \multicolumn{2}{c}{7} & \multicolumn{2}{c}{7} &\multicolumn{2}{c}{6}\\
\hline
\multicolumn{16}{c}{Patches for Downstream Task}\\
\hline
\multirow{2}{*}{Dataset}&\multicolumn{3}{c|}{Train} & \multicolumn{6}{c|}{Valid} & \multicolumn{6}{c}{Test}\\
&Tissue&Other&TIL&\multicolumn{2}{c}{Tissue}&\multicolumn{2}{c}{Other}&\multicolumn{2}{c|}{TIL}&\multicolumn{2}{c}{Tissue}&\multicolumn{2}{c}{Other}&\multicolumn{2}{c}{TIL}\\
\hline
CoNSeP&300&949&349&\multicolumn{2}{c}{40}&\multicolumn{2}{c}{40}&\multicolumn{2}{c|}{40}&\multicolumn{2}{c}{70}&\multicolumn{2}{c}{491}&\multicolumn{2}{c}{268}\\
\hline
\end{tabular}
\end{table}

\begin{figure}
    \centering
    \includegraphics[width=1.0\textwidth]{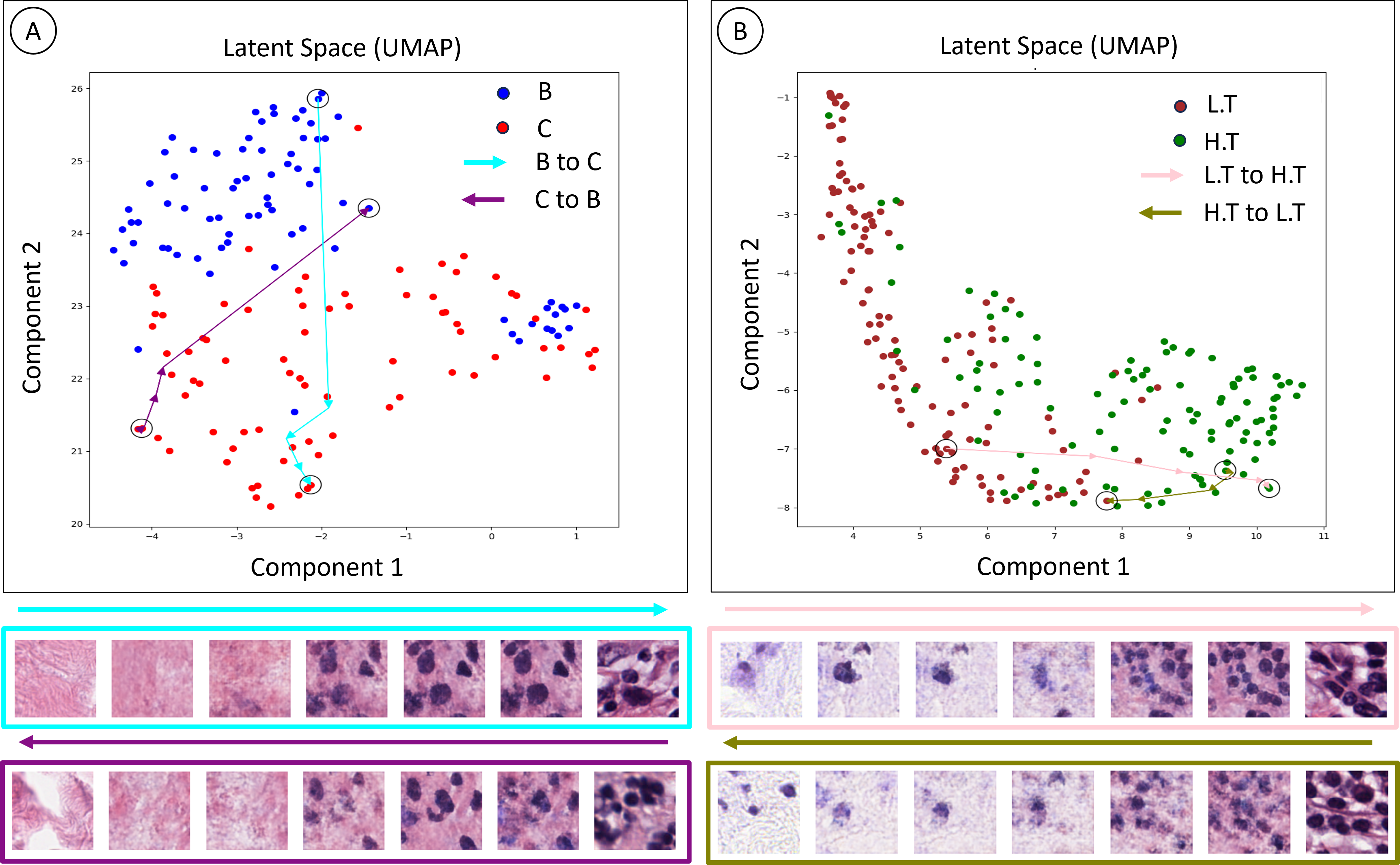}
    \caption{Latent Space Interpolation. A) Salient latent interpolation from one of the samples from $B$ to $C$ and vice versa visualized with UMAP. B) Salient latent interpolation from one of the samples from $L.T$ to $H.T$ and vice versa. The bottom of both (A) and (B) represents reconstructed images from the interpolated latent. Leftmost and rightmost are original samples.} 
    \label{Latent Space Interpolation}
\end{figure}